\documentclass[structabstract]{aa}
\usepackage{graphicx}
\usepackage{txfonts}
\usepackage{verbatim}
\usepackage{natbib}

\begin{document}

\title{Analysing the active longitudes of the young solar analogue \object{HD 116956} using Bayesian statistics}
\author{J. Lehtinen \inst{1}
\and L. Jetsu \inst{1}
\and M. J. Mantere \inst{1}}
\institute{Department of Physics, Gustaf H\"{a}llstr\"{o}min katu 2a (P.O. Box 64), FI-00014 University of Helsinki, Finland}
\date{Received / Accepted}

\abstract{}
{In this study, we aim at investigating the properties of the active longitude system of the young solar analogue \object{HD 116956} in detail, especially concentrating on determining the rotation period of the spot-generating mechanism with respect to the photometric rotation period of the star itself. Because the nonparametric approach, like the Kuiper method, can only give the period of active longitudes, we formulate a new method that can determine the parameters the active longitude distribution uniquely.}
{For this purpose, we have developed an analysis method, based on Bayesian statistics using Markov chain Monte Carlo, presented in this manuscript. One of the advantages of this method is that an estimate of the active longitude system rotation period, as well as the parameters of the shape and location of the active longitudes together with their respective error estimates. This allows us to compare the active longitude and mean photospheric rotation periods of the star.}
{Our analysis confirms previous results of the object having two stable active longitudes with a phase difference of $\Delta \phi \approx 0.5$, the other longitude having dominated over the other one during almost the entire span of the time series. Our method gives the rotation period of the active longitude system $P_{\rm al} = 7.8412 \pm 0.00002$d, which is significantly different from the mean photospheric rotation period of the star $P_{\rm rot} = 7.817 \pm 0.003$d.}
{Our analysis indicates that the spot-generating mechanism, manifesting itself as a system of two active longitudes, is lagging behind the overall rotation of the star. This behaviour may be interpreted as a nonaxisymmetric dynamo wave propagating in the rotational reference frame of the stellar surface.}

\maketitle

\section{Introduction}

In our recent paper \citep{lehtinen2011continuous}, we presented an analysis of 12 years (from late 1998 till mid 2010) of Johnson V band photometry from the star \object{HD 116956}. This star is a chromospherically active young solar analogue having a spectral type of G9V, and an age of the order of a few 100 Myr \citep{gaidos2000spectroscopy}. The spot activity of the star causes rotational modulation of brightness, occasionally reaching as high as $0.07$mag.

Our analysis revealed that the star has stable active longitudes that have retained their identities over the whole 12 years of observations. For the most of this time, one of these active longitudes has been dominant, accompanied by a weaker secondary that seems to disappear and reappear irregularly. The only exception to this behaviour was observed during the first observing season in 1998--1999, where we identified a flip-flop, i.e. switch of the strongest spot activity from one active longitude to the other. The phase separation of the two active longitudes has remained close to $\Delta\phi\approx0.5$. We determined the rotation period of this active longitude structure to be $P_{\rm al}=7.8416\pm0.0011$d, using the formulation of the Kuiper test given by \citet{jetsu1996searching}. This value was somewhat different from the weighted mean photometric rotation period $P_{\rm rot}=7.8288$d suggesting that the active longitudes rotate with a different period than the star itself.

Active longitudes migrating in the rotational reference frame of the star have been found by the spectroscopic study by \citet{lindborg2011doppler} from the \object{RS CVn} binary \object{II Peg}. In this case the active longitudes rotate faster than the orbital period of the star in the synchronised binary system. This is in agreement with the photometric analysis of \citet{berdyugina1998permanent} and \citet{jetsu1996active}. In the photometric studies, migratory active longitudes were found for several other stars, including \object{EI Eri} and \object{HR 7275} \citep{berdyugina1998permanent} and \object{$\lambda$ And} and \object{V711 Tau} \citep{jetsu1996active}. For the stars \object{$\sigma$ Gem} \citep{berdyugina1998permanent} and \object{FK Com} \citep{JPT93}, the two periods seem to coincide. Especially in the case of \object{FK Com}, indications of the migration period being variable in time is piling up based on spectroscopic studies \citep{KBHIST07}.

In this paper, we examine the behaviour of the active longitudes of \object{HD 116956} in greater detail. For this purpose, we formulate a new analysis method based on Bayesian statistics. In Sects. \ref{background} and \ref{setup}, we discuss the background of the statistical analysis of active longitudes and the setup of the probability model used in the Bayesian analysis. In Sect. \ref{modelling}, we formulate the analysis algorithm using Markov chain Monte Carlo. This algorithm is then applied to the data from \object{HD 116956} in Sect. \ref{analysis}. Finally, in Sect. \ref{dynamo}, we discuss our results in the light of dynamo theory.

\section{Background}
\label{background}

For the analysis of active longitudes, we need position information of the active areas on the star as a function of time. Minimal necessary data consists of central meridian passing times $t_i$ of the active areas. In the case of photometric time series analysis, these are provided by the light curve minimum epochs. If the active areas on the star are concentrated on stable active longitudes, the $t_i$ should occur regularly with the active longitude rotation period $P_{\rm al}$. Longitudinal position information, in the form of phases $\phi_i$ of the inferred active areas, is obtained by folding the $t_i$ with the period $P_{\rm al}$. It is typically well justified to assume the number of active longitudes to be between zero and two. This is because from the photometry alone it is very difficult to separate two simultaneously active areas from each other if their phase separation is smaller than $\phi\approx0.33$ \citep[see][]{lehtinen2011continuous}.

To determine the value of $P_{\rm al}$, the minimum times $t_i$ are typically analysed with some nonparametric analysis method, such as the Kuiper test \citep{kuiper1960tests,jetsu1996searching}. Such an analysis was done by \citet{jetsu1996active} for four active stars to find active longitudes in their photometry. A more detailed quantitative analysis can be built on the basis of a preliminary nonparametric search for $P_{\rm al}$ by parametrising the full phase distribution of $\phi_i$. This way we can both improve the precision with which $P_{\rm al}$ is determined, but also quantify the shape and location of the active longitudes in the reference frame rotating with the period $P_{\rm al}$. The solution is obtained by using the tools of Bayesian statistics.

In this paper, we analyse the light curve minimum epochs of \object{HD 116956} determined with the CPS method \citep{lehtinen2011continuous}. The input data consists of all reliable primary and secondary minimum epochs $t_{1,i}$ and $t_{2,i}$, and their error estimates $\sigma_{t_{1,i}}$ and $\sigma_{t_{2,i}}$. These epochs are given in Heliocentric Julian Dates.

\section{Setup of the model}
\label{setup}

We use Bayesian inference for modelling the active longitudes. In other words, we compute the joint posterior probability density $p(\theta|y)$ of the model parameters $\theta$ conditional to the observed data $y$ using the Bayes' formula \citep{gelman2004bayesian}
\begin{equation}
p(\theta|y) = \frac{p(\theta)p(y|\theta)}{p(y)}.
\label{post}
\end{equation}

The important constituents in the computation of the posterior density are the prior density $p(\theta)$ and the likelihood function $p(y|\theta)$. The likelihood function is the actual probability model for the data. It gives the probability density of the observed data conditional to given values of the model parameters. The prior density $p(\theta)$ quantifies any existing information about the model parameters. This can be, for example, information gathered during previous experiments. The prior can also be formulated so that it quantifies the lack of information about the parameter values in an objective way. The factor $p(y)=\int{p(\theta)p(y|\theta)}{\rm d}\theta$ depends only on the observed data and acts as the normalisation coefficient of the posterior density $p(\theta|y)$.

Knowing the full probability distribution for each of the model parameters, i.e. the marginal posterior densities, we can calculate the parameter estimates as means or medians, and their errors as standard deviations or posterior probability intervals of the respective distributions.

\subsection{Likelihood function}

We are interested in modelling the distribution of phases of the light curve minima. Thus, our likelihood function $p(y|\theta)$ should be some circular probability distribution. Let us assume, that there is an active longitude on the stellar surface producing observable active areas around some phase $\phi_0$ and that there is both physical and observational uncertainty in this phase. It is natural to think that the phase uncertainty has a Gaussian distribution. Using phase angle $\psi=2\pi\phi$ ($\psi\in[0,2\pi]$) this is equivalent to the wrapped normal distribution \citep{batschelet1981circular},
\begin{eqnarray*}
p(\psi|\psi_0,\sigma) = \frac{1}{\sigma\sqrt{2\pi}}\sum_{k=-\infty}^{\infty}{\rm e}^{-(\psi-\psi_0+2\pi k)^2/2\sigma^2}.
\end{eqnarray*}

Unfortunately the wrapped normal distribution cannot be written in a closed form. A standard approximation is the von Mises distribution \citep{batschelet1981circular},
\begin{eqnarray*}
p(\psi|\psi_0,\kappa) = \frac{1}{2\pi I_0(\kappa)}{\rm e}^{\kappa\cos{(\psi-\psi_0)}}
\end{eqnarray*}
for phase angle $\psi\in[0,2\pi]$. Here $I_0(x)$ denotes the modified Bessel function of the order 0. For the concentration parameter $\kappa=0$, this gives a uniform distribution in phase angle and for large $\kappa$ this approaches the wrapped normal distribution with $\sigma^2=\kappa^{-1}$.

We are interested in analysing our observations using phases $\phi$ rather than phase angles $\psi$. Therefore we rewrite the von Mises distribution as
\begin{equation}
p(\phi|\phi_0,\kappa) = \frac{1}{I_0(\kappa)}{\rm e}^{\kappa\cos{(2\pi(\phi-\phi_0))}}
\label{vm}
\end{equation}
for phase $\phi\in[0,1]$.

One von Mises distribution is adequate for describing one active longitude. However, there is often reason to suspect that the star being analysed has two active longitudes rather than one. These can be modelled by a mixture of two von Mises distributions as
\begin{equation}
p(\phi|\theta) = m\frac{{\rm e}^{\kappa\cos{(2\pi(\phi-\phi_{0,1}))}}}{I_0(\kappa_1)} + (1-m)\frac{{\rm e}^{\kappa\cos{(2\pi(\phi-\phi_{0,2}))}}.}{I_0(\kappa_2)},
\label{like}
\end{equation}
where $\theta=[m,\phi_{0,1},\phi_{0,2},\kappa_1,\kappa_2]$ are the free parameters of the mixture model. To ensure correct normalisation, the mixture parameter $m$ has to be restricted to the interval $m\in[0,1]$.

As the data in reality consists of $n$ data points, the actual likelihood function is a product of likelihoods for each of these,
\begin{equation}
p(\phi|\theta) = \prod_{i=1}^n{p(\phi_i|\theta)}.
\label{liken}
\end{equation}
It is also possible to use weights for the individual data points, in which case the likelihood becomes
\begin{equation}
p_{\rm w}(\phi|\theta) = \prod_{i=1}^n{p(\phi_i|\theta)^{w_i}}.
\label{wlik}
\end{equation}
Here the normalised weights $w_i$ can be obtained from any relative
weights $\tilde{w}_i$ by dividing with their mean
\begin{equation}
w_i = \frac{n\tilde{w}_i}{\sum_{j=1}^n{\tilde{w}_j}}.
\end{equation}
Hence, the mean of $w_i$ is 1. This normalisation of the weights is important, because the product in Eq. \ref{wlik} must contain the same amount of information from the $n$ available data points as the nonweighted product in Eq. \ref{liken}.

Lastly, to be able to model the active longitudes from the observed light curve minimum epochs, we need the active longitude period $P_{\rm al}$ (hereafter called $P$ when used to denote the free model parameter) to convert these time points $t_i$ into phases $\phi_i$. This is done by
\begin{equation}
\phi_i(P) = FRAC\left[\frac{t_i}{P}\right],
\label{frac}
\end{equation}
where $FRAC[x]$ removes the integer part of $x$. The period $P$ need not to be given, but can be considered as the sixth free model parameter together with the other free parameters $\theta$ of Eq. \ref{like}. The period $P$ is included in the likelihood function implicitly via Eq. \ref{frac}.

\subsection{Parameter priors}

In order to compute the posterior density $p(\theta|y)$, we need priors for the model parameters $\theta$. We choose noninformative priors, i.e. priors that quantify our lack of knowledge about the values of these free parameters.

For the active longitude phases $\phi_{0,1}$ and $\phi_{0,2}$, the most natural choice for a noninformative prior is a uniform prior in the interval $[0,1]$. This states that \textit{a priori} we consider that the active longitudes may lie at any phase with equal probability. Similarly, the mixture parameter $m$ is bound to the interval $[0,1]$ and is naturally given the same uniform prior. Note, however, that the priors $p(\phi_{0,1})$ and $p(\phi_{0,2})$ are only formally bound between 0 and 1. In actual modelling, we allow $\phi_{0,1}$ and $\phi_{0,2}$ to get values below 0 as well as above 1. The sole purpose of this approach is to take care that the wings of the marginal posterior densities of $\phi_{0,1}$ and $\phi_{0,2}$ would not be cut at the phases $\phi=0$ or $1$.

For the concentration parameters $\kappa_1$ and $\kappa_2$, we choose the improper, i.e. non-normalisable, Jeffreys' prior \citep{dobigeon2007joint}
\begin{equation}
p(\kappa)=\frac{1}{\kappa},
\end{equation}
for $\kappa>0$. This is constructed specially to reflect vague information about the value of $\kappa$. The same prior works also for the period \citep{gregory1992new}, i.e.
\begin{equation}
p(P)=\frac{1}{P},
\label{prip}
\end{equation}
for $P>0$. The fact that this prior is noninformative about $P$ can be seen by a transformation of variables $f=P^{-1}$ from the period domain to the frequency domain. Under this change, the prior given by Eq. \ref{prip} retains its shape, just as a noninformative prior should.

The joint prior of all the model parameters is just the product of all the individual parameter priors,
\begin{eqnarray*}
p(\theta) = p(P)p(m)p(\phi_{0,1})p(\phi_{0,2})p(\kappa_1)p(\kappa_2).
\end{eqnarray*}

\section{Modelling algorithm}
\label{modelling}

To estimate the joint posterior density $p(\theta|y)$ of the model parameters, we need to use a Monte Carlo approach. In other words, we simulate random numbers that are distributed according to $p(\theta|y)$. This is possible by using the Markov chain Monte Carlo (hereafter MCMC) approach \citep{gilks1996introducing}.

\subsection{Preliminary estimation of $P$}

To ensure that the MCMC algorithm indeed converges to the correct parameter values, it is necessary to make a preliminary search first for the posterior mode within broad search intervals. Because our model (Eqs. \ref{like}--\ref{frac}) is rather complex, this search is done in several stages.

The first model parameter to be searched for is the active longitude period $P$. For this search we use the Kuiper periodogram \citep{kuiper1960tests,jetsu1996searching}, constructed by computing the value of the Kuiper test statistic $V_n(f)$ for a discrete grid of frequency values $f=P^{-1}$.

Computing the Kuiper test statistic consists of comparing the empirical cumulative distribution function $F_n(\phi)$ of the phases $\phi_i$ to some theoretical cumulative distribution function $F(\phi)$. As we are interested in finding any deviations from uniform distribution of $\phi_i$, the theoretical cumulative distribution function in our case is $F(\phi)=\phi$. The test statistic is
\begin{eqnarray*}
V_n(f) = D^+ + D^-,
\end{eqnarray*}
where $D^+$ and $D^-$ denote the maximum values of $F_n(\phi)-F(\phi)$ and $F(\phi)-F_n(\phi)$, respectively.

As stated above, the Kuiper periodogram is computed for a fixed set of frequency values. This is most naturally done with a grid centered around the initial guess for the frequency $f_0=P_0^{-1}$. The tested frequencies are within the range
\begin{eqnarray*}
(1-q)f_0 \leq f \leq (1+q)f_0,
\end{eqnarray*}
and are separated by evenly spaced steps of 
\begin{eqnarray*}
\Delta f = [(t_n-t_1)OFAC]^{-1}
\end{eqnarray*}
from each other, where $t_1$ and $t_n$ are the first and the last time point, respectively, and $OFAC$ is the over filling factor. In this paper, we have used $q=0.05$ and $OFAC=10$. The period value $P=f^{-1}$ which maximises the periodogram, i.e. gives the largest value of $V_n(f)$, is used as the preliminary estimate for $P$.

\subsection{Preliminary estimation of $m$, $\phi_{0,1}$, $\phi_{0,2}$, $\kappa_1$ and $\kappa_2$}

With a preliminary estimate for $P$, we can start searching for estimates for the rest of the model parameters. This can be consistently done with the full MCMC strategy, by finding the approximate mode of the posterior density (Eq. \ref{post}) conditional to the value obtained for $P$.

To speed up the process of finding the approximate posterior mode, it is useful to divide the parameters into two groups. We first search for the modes of $m$, $\phi_{0,1}$ and $\phi_{0,2}$, each within the interval $[0,1]$. This is done by computing the value of the posterior density with proposed values of $m$, $\phi_{0,1}$ and $\phi_{0,2}$, using preliminary estimate for $P$ and some reasonable initial guesses for $\kappa_1$ and $\kappa_2$, and taking steps towards higher posterior density. Once reasonable estimates for $m$, $\phi_{0,1}$ and $\phi_{0,2}$ have been obtained, we search for the modes of $\kappa_1$ and $\kappa_2$. This is done by computing the posterior density for a grid of $\kappa_1$ and $\kappa_2$ conditional to the preliminary estimates of the rest of the parameters. In this paper, we do the initial search between $\kappa_{\rm min}=0.5$ and $\kappa_{\rm max}=50.0$ and with a grid spacing of $\Delta\kappa=0.1$.

\subsection{Posterior simulation}

After crude estimates of the model parameters have been obtained, the full posterior density $p(\theta|y)$ is modelled with a MCMC algorithm \citep{gilks1996introducing}. This algorithm generates random values for the model parameters as Markov chains. In other words, we simulate $n_{\rm MCMC}$ values for all of the model parameters $\theta$, so that the values on each simulation round, $\theta_t$, depend on their values on the previous round, $\theta_{t-1}$. When the algorithm is executed correctly, these chains converge to the marginal posterior densities of the individual model parameters. The preliminary parameter estimates are used as the starting values for the Markov chains.

Because of the complex nonlinearity of the active longitude model (Eqs. \ref{like}--\ref{frac}), we split the MCMC algorithm into several stages. During these, a limited number of free parameter values are updated.

The first parameter to be updated is the mixture parameter $m$. For this, we compute the value of the posterior density at each minimum phase with the parameter values of the last simulation round and assuming likelihoods containing only one active longitude, i.e. either $p(\phi_i|1)=p(\phi_i|\phi_{0,1},\kappa_1)$ or $p(\phi_i|2)=p(\phi_i|\phi_{0,2},\kappa_2)$. Using these, we compute the probability of $\phi_i$ belonging to the active longitude 1,
\begin{equation}
z_i = \frac{m_{t-1}p(\phi_i|1)}{m_{t-1}p(\phi_i|1) + (1-m_{t-1})p(\phi_i|2)},
\end{equation}
where $m_{t-1}$ denotes the value of the mixture parameter on the previous simulation round. The new value for $m$ is now given by the weighted mean of $z_i$ for all $\phi_i$,
\begin{equation}
m_t = \frac{\sum_{i=1}^n{w_iz_i}}{\sum_{i=1}^n{w_i}}.
\end{equation}
This procedure for simulating $m$ is based on the strategy proposed by \citet{gelman2004bayesian} for computing mixture models.

The computation of the probabilities $z_i$ also allows us to select the active longitude that the minimum phases $\phi_i$ belong to. The division is done at each simulation round and is probabilistic, i.e. each $\phi_i$ is assigned to the active longitude 1 with the probability $z_i$. Those minimum phases that are not assigned to the first active longitude are automatically assigned to the second active longitude. This division of the minimum phases makes it possible to model both of the active longitudes separately.

After a new value for $m$ has been simulated, we update the values of the period $P$ and the phases of both of the active longitudes, $\phi_{0,1}$, $\phi_{0,2}$, using the bimodal likelihood function (Eq. \ref{like}) with the new value for $m$ and old values for $\kappa_1$ and $\kappa_2$. The simulation is done with a standard Metropolis-Hastings algorithm \citep{gilks1996introducing}, where new proposed values $\theta_t$ for $m$, $\phi_{0,1}$ and $\phi_{0,2}$ are drawn from a Gaussian proposition density $q(\theta_t)=N(\theta_t|\theta_{t-1},\sigma_{q,\theta}^2)$ centered around the parameter value $\theta_{t-1}$ of the previous simulation round and having a different adjustable variance $\sigma_{q,\theta}^2$ for the different parameters.

Last, we update the values of the concentration parameters $\kappa_1$ and $\kappa_2$. Since obtaining good mixing for the Markov chains of these parameters can be tricky, we do the updating separately for both of these parameters using a unimodal likelihood function (Eq. \ref{vm}) and only the minimum phases labelled as those belonging to the active longitude of the currently updated concentration parameter.

We use a gamma distribution as the proposition density for the concentration parameters, $q(\kappa_{t})={\rm Gamma}(\kappa_{t}|\alpha,\beta)$, as suggested by \citet{dobigeon2007joint}. The parameters for this distribution are set to be $\alpha=\hat{\kappa}^2/\sigma_{q,\kappa}^2$ and $\beta=\hat{\kappa}/\sigma_{q,\kappa}^2$, where $\hat{\kappa}$ denotes the preliminary estimates for the concentration parameters and $\sigma_{q,\kappa}^2$ is an adjustable variance for the proposition density.

It usually takes some amount of simulation rounds before the Markov chains converge to the marginal posterior densities of the model parameters. To eliminate this burn-in phase in the beginning of the simulation, it is necessary to discard the first $n_{\rm burn-in}$ simulation rounds from the chains. In short simulations, it may be necessary to set $n_{\rm burn-in}$ to be 50 \% of the total length of the simulation. In longer simulations, 10 \% of the total length is often sufficient.

The marginal posterior densities of the model parameters might not be very illustrative as themselves. Therefore we summarise the results of the MCMC simulation by computing estimates for the parameter values and their errors. If the parameters turn out to have nearly Gaussian distributions, it is natural to estimate their values with the means and their errors with their standard deviations of the respective Markov chains. For more complicated distributions, it is better to give the median and some quantiles $q$ and $1-q$ of the Markov chains as the parameter and error estimates.

\section{Analysis of the active longitudes}
\label{analysis}

We use the MCMC algorithm described above to analyse the active longitudes of \object{HD 116956}. The original observations were made between $\rm HJD=2451172$ (24th of December, 1998) and $\rm HJD=2455342$ (25th of May, 2010) with the T3 0.4 m automatic photoelectric telescope (APT) at the Fairborn Observatory in Arizona. They were divided into 12 continuous observing seasons, each spanning roughly the first half of the year. We labelled these observing seasons as segments SEG 1--12 \citep{lehtinen2011continuous}. The CPS method  provided estimates for both the primary and secondary minimum epochs, $t_{1,i}$ and $t_{2,i}$. However, in our analysis of this paper we pool these two classes together as $t_i$, where the number of time points is $n=645$. This pooling is done because, even if the star has two stable active longitudes, both the primary and secondary light curve minima can very well appear on both of the active longitudes. This is possible if the levels of spot activity on the two active longitudes are close to each other and either one of them may be stronger at any given moment. Another possible alternative is a flip-flop, where the main spot activity physically shifts from one active longitude to the other.

For the analysis in this paper, we use weights. In other words, we associate a weight $w_i$ for each data point $t_i$ and use Eq. \ref{wlik} to combine the likelihoods of these data points. A typical definition for relative weights is the inverse square of error, i.e. $\tilde{w}_i=\sigma_{t_i}^{-2}$. This is also what we use in the CPS. It may, however, occur that the weights obtained this way have a large dispersion and some of them have values greatly larger than others. In this case, the burden of carrying information from the data to the posterior density is strongly concentrated on just a few data points. This can in turn lead to an instability in the posterior simulation. To even out the pathologically defined weights, we must use a more robust definition for them, e.g. $\tilde{w}_i=\sigma_{t_i}^{-1}$. We use this definition for the weights throughout the paper.

\subsection{Global and seasonal analyses}
\label{alan}

\begin{figure}
\resizebox{\hsize}{!}{\includegraphics{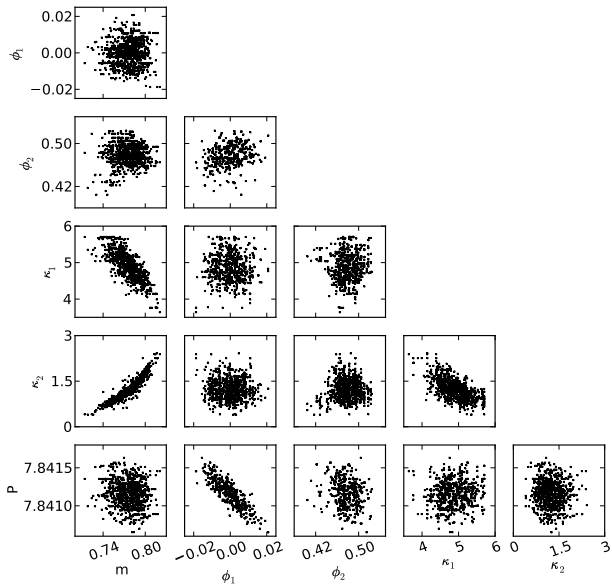}}
\caption{The simulated parameter values of the MCMC run for the global analysis. Presented are scatter plots between each pair of model parameters showing the correlations between the parameters.}
\label{corr}
\end{figure}

\begin{table}
\caption{Parameter and error estimates for global analysis of the active longitudes. The second column gives the means and standard deviations of the Markov chain of $P$, $\phi_1$ and $\phi_2$. The third and fourth columns give the medians and 5th and 95th centilles of the Markov chains of all of the parameters. The active longitude rotation period $P$ is given in days while the rest of the parameters are dimensionless.}
\center
\begin{tabular}{l l l c}
\hline
\hline
& mean $\pm$ std & median & [5\%,\ \ 95\%] \\
\hline
$P_{\rm al} \ [{\rm d}]$  & $7.8412\pm0.0002$ & 7.8412  & [7.8409,\ \ 7.8414] \\
$m$        & \ldots            & 0.78    & [0.75,\ \ 0.80] \\
$\phi_1$   & $0.0\pm0.007$     & 0.0     & [-0.011,\ \ 0.011] \\
$\phi_2$   & $0.479\pm0.019$   & 0.479   & [0.444,\ \ 0.511] \\
$\kappa_1$ & \ldots            & 4.88    & [4.28,\ \ 5.62] \\
$\kappa_2$ & \ldots            & 1.21    & [0.75,\ \ 1.86] \\
\hline
\end{tabular}
\label{globtab}
\end{table}

We are interested in both the long term overall behaviour, as well as any seasonal changes of the active longitudes. Thus, we present both a global analysis, including all the available minimum epochs, and seasonal analyses that include minimum epochs only from one observing segment at a time. For each case, we ran the MCMC algorithm with $n_{\rm MCMC}=10000$ rounds and discarded the first $n_{\rm burn-in}=1000$ as the burn-in phase.

An idea of how the MCMC algorithm works in practice, is visualised in Fig. \ref{corr}. It shows the Markov chains of the model parameters $\theta$ in the global analysis as scatter plots between pairs of free parameters. As the burn-in phase has been discarded, these scatter plots should reflect the shape of the multidimensional joint posterior density. In an ideal case, each panel in Fig. \ref{corr} should approximate a bivariate Gaussian with vanishing correlation between the parameters. In reality, there are some parameter pairs, such as $m$ and $\kappa_2$, which show relatively strong correlation or anticorrelation. This shows that the parametrisation of the problem is not the best possible. It is, however, not clear what kind of a reparametrisation could remove this correlation. In any case, the present more or less "physical" parametrisation seems also to yield good mixing of the Markov chains, i.e. the joint posterior density is uniformly sampled. In addition to significant correlation, some panels of Fig. \ref{corr} show slightly irregular shapes. Thus, the joint posterior density has a more complex form than what could have been expected \textit{a priori}.

Parameter estimates of the global analysis are given in Table \ref{globtab}. Because the marginal posterior densities of $m$, $\kappa_1$ and $\kappa_2$ are asymmetric, i.e. not Gaussian, we estimate the parameter values with the medians of the Markov chains of each parameter. For error estimates, we give the $[5\%,95\%]$ posterior intervals. The marginal posterior densities of $\phi_1$, $\phi_2$ and $P$, now denoted with $P_{\rm al}=P$ being regarded as the rotation period of the active longitude system, have more closely Gaussian forms, and for them, we also give the mean and standard deviation of the Markov chains. The two estimates based on the mean or median are in good agreement with each other.

\begin{figure}
\resizebox{\hsize}{!}{\includegraphics{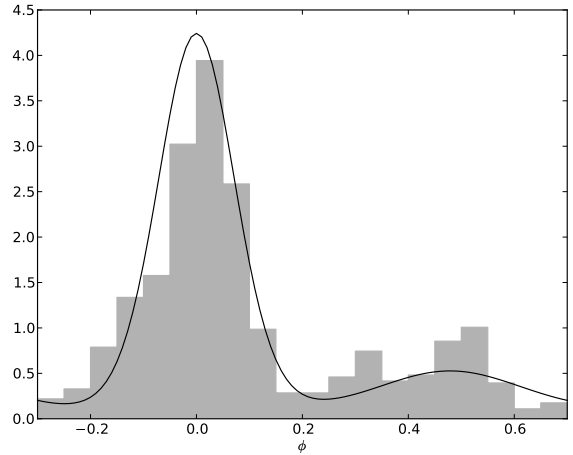}}
\caption{The distribution of all light curve minima folded with the ephemeris ${\rm HJD}_{\rm al}=2451177.6135+7.4812E$ (Eq. \ref{ephem}) and presented as a histogram between the phase interval $-0.5<\phi<0.5$. On top of the observational distribution, we show the active longitude model distribution using the estimated model parameters from Table \ref{globtab}.}
\label{hist}
\end{figure}

Based on the global analysis, we calculated a new linear ephemeris for the central meridian passing epochs of the primary active longitude (labelled with 1),
\begin{equation}
{\rm HJD}_{\rm al}=2451177.6135+7.4812E,
\label{ephem}
\end{equation}
where $E$ is the number of rotations. The model distribution (Eqs. \ref{like} and \ref{frac}) using the estimated parameter values is compared with the observational distribution of the light curve minima in Fig. \ref{hist} folded with the period $P_{\rm al}=7.4812$ d. As suggested by the value $m=0.78$, the primary active longitude is clearly the dominant structure on the star with the secondary active longitude only barely visible. The phase difference between the two active longitudes is quite close to $\Delta\phi=0.5$. The model fit to the observational phase distribution is reasonably good, although not perfect. The differences may be due to seasonal changes in the active longitude phases.

\begin{table}
\caption{Parameter and error estimates for local segment by segment analyses of the active longitudes presented in the same form as in Table \ref{globtab}. The phase difference $\Delta\phi$ between the primary and following secondary active longitude is also given, when both of the longitudes are present. The active longitude period of the global analysis, $P_{\rm al}=7.8412$ d, has been used for all of the segments. The year during which the largest part of the segment data was observed is also given.}
\center
\begin{tabular}{l l l l c}
\hline
\hline
& & mean $\pm$ std & median & [5\%,\ \ 95\%] \\
\hline
SEG 1 & $m$ & \ldots & 0.67 & [0.65,\ \ 0.67] \\
(1999) & $\phi_1$ & $0.966\pm0.008$ & 0.966 & [0.952,\ \ 0.979] \\
& $\phi_2$ & $0.352\pm0.010$ & 0.353 & [0.335,\ \ 0.368] \\
& $\Delta\phi$ & $0.386\pm0.012$ & 0.386 & [0.367,\ \ 0.407] \\
& $\kappa_1$ & \ldots & 5.37 & [4.07,\ \ 6.98] \\
& $\kappa_2$ & \ldots & 6.67 & [4.05,\ \ 10.03] \\
\hline
SEG 2 & $m$ & \ldots & 0.75 & [0.55,\ \ 0.85] \\
(2000) & $\phi_1$ & $0.929\pm0.018$ & 0.929 & [0.900,\ \ 0.959] \\
& $\phi_2$ & $0.211\pm0.056$ & 0.219 & [0.106,\ \ 0.290] \\
& $\Delta\phi$ & $0.282\pm0.047$ & 0.289 & [0.194,\ \ 0.348] \\
& $\kappa_1$ & \ldots & 4.01 & [2.59,\ \ 7.26] \\
& $\kappa_2$ & \ldots & 3.09 & [1.23,\ \ 6.85] \\
\hline
SEG 3 & $m$ & \ldots & 0.62 & [0.62,\ \ 0.62] \\
(2001) & $\phi_1$ & $0.052\pm0.009$ & 0.052 & [0.038,\ \ 0.066] \\
& $\phi_2$ & $0.517\pm0.012$ & 0.516 & [0.497,\ \ 0.538] \\
& $\Delta\phi$ & $0.465\pm0.012$ & 0.465 & [0.446,\ \ 0.485] \\
& $\kappa_1$ & \ldots & 33.46 & [22.35,\ \ 47.24] \\
& $\kappa_2$ & \ldots & 17.51 & [10.21,\ \ 27.27] \\
\hline
SEG 4 & $m$ & \ldots & 0.81 & [0.80,\ \ 0.81] \\
(2002) & $\phi_1$ & $0.046\pm0.015$ & 0.047 & [0.020,\ \ 0.068] \\
& $\phi_2$ & $0.580\pm0.034$ & 0.582 & [0.521,\ \ 0.634] \\
& $\Delta\phi$ & $0.534\pm0.034$ & 0.534 & [0.476,\ \ 0.589] \\
& $\kappa_1$ & \ldots & 9.21 & [6.52,\ \ 12.58] \\
& $\kappa_2$ & \ldots & 4.24 & [1.81,\ \ 7.85] \\
\hline
SEG 5 & $m$ & \ldots & 1.00 & [1.00,\ \ 1.00] \\
(2003) & $\phi_1$ & $0.069\pm0.019$ & 0.069 & [0.037,\ \ 0.103] \\
& $\kappa_1$ & \ldots & 4.95 & [3.64,\ \ 6.50] \\
\hline
SEG 6 & $m$ & \ldots & 1.00 & [1.00,\ \ 1.00] \\
(2004) & $\phi_1$ & $0.025\pm0.021$ & 0.026 & [-0.012,\ \ 0.058] \\
& $\kappa_1$ & \ldots & 15.52 & [7.76,\ \ 23.86] \\
\hline
SEG 7 & $m$ & \ldots & 0.81 & [0.81,\ \ 0.81] \\
(2005) & $\phi_1$ & $0.816\pm0.024$ & 0.816 & [0.778,\ \ 0.851] \\
& $\phi_2$ & $0.443\pm0.025$ & 0.445 & [0.400,\ \ 0.479] \\
& $\Delta\phi$ & $0.627\pm0.025$ & 0.627 & [0.586,\ \ 0.663] \\
& $\kappa_1$ & \ldots & 5.22 & [3.21,\ \ 7.01] \\
& $\kappa_2$ & \ldots & 30.13 & [6.96,\ \ 53.36] \\
\hline
SEG 8 & $m$ & \ldots & 1.00 & [1.00,\ \ 1.00] \\
(2006) & $\phi_1$ & $0.976\pm0.026$ & 0.977 & [0.932,\ \ 1.019] \\
& $\kappa_1$ & \ldots & 6.14 & [3.55,\ \ 8.47] \\
\hline
SEG 9 & $m$ & \ldots & 0.81 & [0.81,\ \ 0.81] \\
(2007) & $\phi_1$ & $0.012\pm0.029$ & 0.012 & [-0.031,\ \ 0.058] \\
& $\phi_2$ & $0.500\pm0.039$ & 0.502 & [0.438,\ \ 0.566] \\
& $\Delta\phi$ & $0.488\pm0.037$ & 0.490 & [0.428,\ \ 0.548] \\
& $\kappa_1$ & \ldots & 14.80 & [5.31,\ \ 19.56] \\
& $\kappa_2$ & \ldots & 12.75 & [2.78,\ \ 24.62] \\
\hline
SEG 10 & $m$ & \ldots & 1.00 & [1.00,\ \ 1.00] \\
(2008) & $\phi_1$ & $0.005\pm0.033$ & 0.003 & [-0.050,\ \ 0.061] \\
& $\kappa_1$ & \ldots & 2.50 & [1.81,\ \ 3.10] \\
\hline
SEG 11 & $m$ & \ldots & 0.71 & [0.67,\ \ 0.75] \\
(2009) & $\phi_1$ & $0.910\pm0.035$ & 0.910 & [0.852,\ \ 0.970] \\
& $\phi_2$ & $0.315\pm0.043$ & 0.318 & [0.243,\ \ 0.378] \\
& $\Delta\phi$ & $0.405\pm0.038$ & 0.406 & [0.340,\ \ 0.426] \\
& $\kappa_1$ & \ldots & 3.40 & [1.89,\ \ 4.95] \\
& $\kappa_2$ & \ldots & 5.13 & [2.21,\ \ 8.14] \\
\hline
SEG 12 & $m$ & \ldots & 1.00 & [1.00,\ \ 1.00] \\
(2010) & $\phi_1$ & $0.931\pm0.073$ & 0.936 & [0.834,\ \ 1.045] \\
& $\kappa_1$ & \ldots & 1.05 & [0.89,\ \ 1.88] \\
\hline
\end{tabular}
\label{loctab}
\end{table}

\begin{figure}
\resizebox{\hsize}{!}{\includegraphics{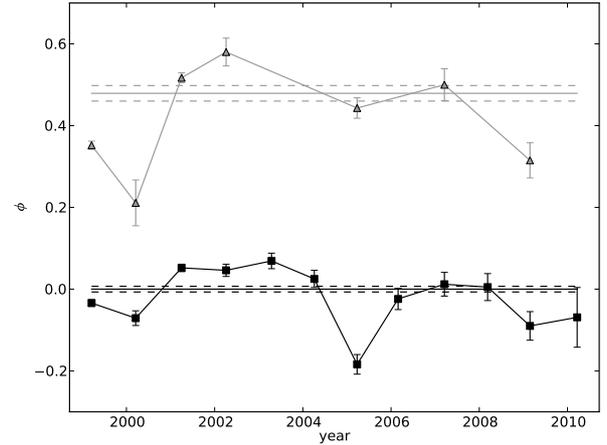}}
\caption{Active longitude phases folded with the ephemeris ${\rm HJD}_{\rm al}=2451177.6135+7.4812E$ for primary (black symbols) and secondary (grey symbols) active longitudes. Individual symbols denote seasonal and horisontal lines global estimates. The $1\sigma$ errors are denoted with error bars for the seasonal and with dashed lines for the global estimates.}
\label{alplot}
\end{figure}

The seasonal changes of the active longitudes were investigated by setting the period $P=P_{\rm al}=7.8412$ d and not letting this value vary in the MCMC runs of the individual segments. Parameter estimates for each of the 12 segments are presented in Table \ref{loctab}. The phases are consistent with the ephemeris of Eq. \ref{ephem}, and therefore the primary active longitude phase $\phi_1$ has values near both 0 and 1. Table \ref{loctab} also includes the phase difference $\Delta\phi$ of the primary active longitude and the following secondary active longitude, when the two have been present.

The seasonal active longitude phases, $\phi_1$ and $\phi_2$, are plotted in Fig. \ref{alplot} along with the phases from the global analysis. It is immediately clear, that the active longitudes have undergone significant shifts. This is in good agreement with our previous results \citep[Fig. 10 in][]{lehtinen2011continuous}. In the present analysis, also the phase difference $\Delta\phi$ between the active longitudes seems to vary significantly. This may, however, be caused by the small number of secondary light curve minima available in many of the segments, giving a less stable simulation of the secondary active longitude phase.

In the segments 5,\ 6,\ 8,\ 10 and 12, the MCMC algorithm yields mixture parameter values of $m=1$, meaning that only the primary active longitude is present. The algorithm does not determine the number of active longitudes present in the data, but uses the bimodal model of Eq. \ref{like} for all data. In spite of this, it was able to detect the absence of the secondary active longitude in these five segments.

Lastly, it is worth noting that the concentration parameters $\kappa_1$ and $\kappa_2$ have very different values in individual segments and typically also wide $[5\%, 95\%]$ posterior intervals within each segment. This demonstrates the fact that these parameters are the least stable ones of our model. The typical problem of borderline data is often just to get decent mixing for the Markov chains of $\kappa_1$ and $\kappa_2$. This is, however, not a major concern, since the exact values of $\kappa_1$ and $\kappa_2$ are of little interest. Our main goal is to determine the values of $P$, $\phi_1$ and $\phi_2$.

\begin{table}
\caption{Seasonal active longitude periods with $1\sigma$ errors.}
\center
\begin{tabular}{l l}
\hline
\hline
SEG & $P_{\rm al}\ {\rm [d]}$ \\
\hline
1 & $7.8871\pm0.0220$ \\
2 & $7.7031\pm0.0136$ \\
3 & $7.8067\pm0.0077$ \\
4 & $7.8952\pm0.0115$ \\
5 & $7.8147\pm0.0134$ \\
6 & $7.8651\pm0.0080$ \\
7 & $7.8752\pm0.0097$ \\
8 & $7.9128\pm0.0046$ \\
9 & $7.8555\pm0.0036$ \\
10 & $7.8193\pm0.0047$ \\
11 & $7.7695\pm0.0131$ \\
12 & $7.8300\pm0.0223$ \\
\hline
\end{tabular}
\label{pertab}
\end{table}

\subsection{Comparison with the mean rotation period}
\label{comp}

\begin{table}
\caption{Global and seasonal weighted mean photometric periods with $1\sigma$ errors.}
\center
\begin{tabular}{l l}
\hline
\hline
SEG & $P_{\rm rot}$ [d] \\
\hline
Global & $7.817\pm0.003$ \\
\hline
1 & $7.890\pm0.014$ \\
2 & $7.698\pm0.014$ \\
3 & $7.808\pm0.010$ \\
4 & $7.803\pm0.021$ \\
5 & $7.801\pm0.009$ \\
6 & $7.855\pm0.010$ \\
7 & $7.830\pm0.011$ \\
8 & $7.836\pm0.008$ \\
9 & $7.817\pm0.005$ \\
10 & $7.807\pm0.010$ \\
11 & $7.839\pm0.017$ \\
12 & $7.659\pm0.044$ \\
\hline
\end{tabular}
\label{phottab}
\end{table}

The global constant value of $P_{\rm al}$ was used for modelling each of the segment in the previous section. We perform another analysis, where the period of each segment is also an additional sixth free parameter, just as in the case of the previous global analysis. The active longitude phases obtained this way are not consistent with each other, but this approach can give information of the changes in the active longitude rotation period. The resulting period estimates presented as means and standard deviations of the Markov chains of $P$ are given in Table \ref{pertab}.

We want to compare the global and seasonal periods of the active longitudes computed in Sect. \ref{alan} with the mean period of the light curve of \object{HD 116956}. The idea is that the photometric periods obtained with the CPS correspond to the rotation of the active areas on the stellar surface. On the other hand, active longitudes that stay stable for a number of years would rather seem to represent the rotation of the underlying global geometry of the magnetic field producing the active areas.

The two kinds of rotation periods need not to coincide. If the rotation period of the magnetic structure significantly differs from the surface rotation period, we would expect to observe the active areas forming at the active longitudes, but then start to drift away.

In \citet{lehtinen2011continuous} we calculated the weighted mean light curve period of \object{HD 116956} from the results of the CPS analysis. To be able to compare the mean period with the active longitude rotation period, we also need an error estimate for the former. Such a weighted mean period and its error estimate can be computed in the Bayesian paradigm consistently with the active longitude analysis presented above. We thus need to formulate the posterior density of the mean photometric period $P_{\rm rot}$.

Each period estimate $P_i$ from the CPS can be assumed to have a Gaussian error distribution with the variance $\sigma_{{\rm P}_i}^2$. This corresponds to the likelihood
\begin{eqnarray*}
p(P_i|P_{\rm rot},\sigma_{{\rm P}_i}^2) = N(P_i|P_{\rm rot},\sigma_{{\rm P}_i}^2) = \frac{1}{\sigma_{{\rm P}_i}\sqrt{2\pi}}{\rm e}^{-(P_{\rm rot}-P_i)^2/2\sigma_{{\rm P}_i}^2}.
\end{eqnarray*}
Combined for all the period estimates, this becomes
\begin{eqnarray*}
p(P|P_{\rm rot},\sigma_{\rm P}^2) &=& \prod_{i=1}^n{N(P_i|P_{\rm rot},\sigma_{{\rm P}_i}^2)} \\
&=& \frac{1}{(2\pi)^{n/2}}\prod_{i=1}^n{\left[\sigma_{{\rm P}_i}^{-1}\right]}{\rm e}^{\sum_{j=1}^n{\left[-(P_{\rm rot}-P_j)^2/2\sigma_{{\rm P}_j}^2\right]}}.
\end{eqnarray*}
By using the prior of Eq. \ref{prip} for $P_{\rm rot}$, the posterior density becomes
\begin{equation}
p(P_{\rm rot}|P,\sigma_{\rm P}^2) \propto p(P_{\rm rot})p(P|P_{\rm rot},\sigma_{\rm P}^2) \propto P_{\rm rot}^{-1}{\rm e}^{\sum_{i=1}^n{\left[-(P_{\rm rot}-P_i)^2/2\sigma_{{\rm P}_i}^2\right]}}.
\label{postpw}
\end{equation}
As each $P_i$ here has a different $\sigma_{{\rm P}_i}^2$, the estimate of $P_{\rm rot}$ using Eq. \ref{postpw} is indeed weighted.

\begin{figure}
\resizebox{\hsize}{!}{\includegraphics{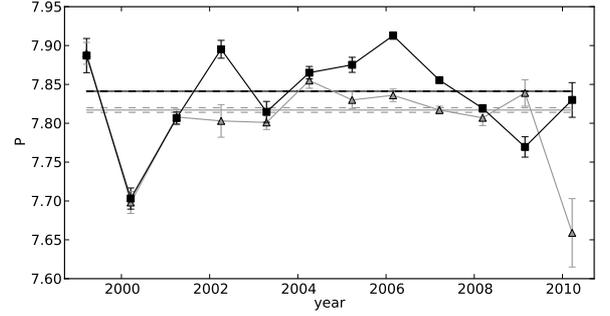}}
\caption{Variations of the seasonal estimates of the active longitude rotation period $P_{\rm al}$ (black squares) and the seasonal photospheric rotation period $P_{\rm rot}$ (grey triangles). The long-term active longitude period from the global analysis and the weighted mean photospheric rotation period are presented with black and grey horisontal lines, respectively. The error bars of the seasonal estimates and dashed lines of the global estimates denote $1\sigma$ error limits.}
\label{perplot}
\end{figure}

Estimates for both the global and seasonal weighted photometric mean periods were obtained by sampling the posterior density of Eq. \ref{postpw} with the Metropolis-Hastings algorithm. For the sampling, we used a Gaussian proposition density $q(P_{{\rm rot},t})=N(P_{{\rm rot},t}|P_{{\rm rot},{t-1}},\sigma_{q,{\rm P}_{\rm rot}}^2)$ with $\sigma_{q,{\rm P}_{\rm rot}}=0.01$ d. For each estimate, we ran the MCMC algorithm with $n_{\rm MCMC}=4000$ rounds, first $n_{\rm burn-in}=2000$ of which were discarded as the burn-in phase. The period estimates presented in Table \ref{phottab} are means of the simulated Markov chains and the respective standard deviations are assigned as their error estimates.

The results are compared graphically with the corresponding active longitude rotation periods in Fig. \ref{perplot}. Both the photometric and active longitude rotation periods seem to have varied quite much during the years. For the first years of observations, the two types of seasonal periods seem to have coincided rather closely, but this behaviour is lost in the more recent observations and the two periods show no mutual correlation. This supports the view that these two periods trace different physical mechanisms on the star.

When comparing the global rotation periods of the active longitudes, $P_{\rm al}=7.8412\pm0.0002$ d, and the photospheric active areas, $P_{\rm rot}=7.817\pm0.003$ d, the difference between the two is clear. The long-term mean photospheric rotation period is significantly shorter than the active longitude rotation period. In the seasonal period estimates, this difference is not that clear because of their large variability. Nevertheless, the effect is visible as all except one seasonal mean photometric periods are shorter than their corresponding active longitude periods or equal within the error limits. Thus it seems, that the active longitudes of \object{HD 116956} rotate more slowly than its photosphere.

\section{Discussion in the light of dynamo theory}
\label{dynamo}

It is generally believed that the magnetic fields in the active rapid rotators are due to a dynamo mechanism working almost solely on the inductive action arising from convective turbulence. The other basic ingredient arising from differential rotation, i.e. the $\Omega$-effect, is expected to be negligibly weak in the rapid rotation regime due to the relative differential rotation $\Delta \Omega/\Omega$ being strongly reduced \citep[e.g.][]{KR99}. Such a dynamo is called as an $\alpha^2$-dynamo, the $\alpha$-effect describing the collective inductive action of turbulence.

In the regime of slower rotation, i.e. in the case of the Sun, the $\Omega$-effect driven by the angular velocity gradients is strong compared to the $\alpha$-effect. For this type of an $\alpha\Omega$ dynamo, oscillating axisymmetric, i.e. field not changing over the azimuthal (longitudinal) coordinate, configurations are typical. The solar nearly dipole field showing a magnetic cycle of roughly 22-years, with a related latitudinal dynamo wave forming the butterfly diagram, is thought to be a representative case of the action of an $\alpha\Omega$-dynamo, see e.g. the review by \citet{O03}.

According to both linear \citep[e.g.][]{krause1980meanfield} and nonlinear solutions \citep[e.g.][]{MBBT95} of the $\alpha^2$-dynamo equations, the nonaxisymmetric modes become more easily excited in the rapid rotation regime. The $m=1$ mode, representing an azimuthally varying field changing sign once over the full longitude span, is commonly the preferred field configuration. This is not surprising, as the largest scale mode is always the most resistive over diffusive effects. The nonaxisymmetric modes turn out to be waves migrating in azimuthal direction, not necessarily having the rotation period of the star \citep[e.g.][]{krause1980meanfield}. Both slower and faster dynamo waves can occur, depending for example on the profile and properties of the turbulent transport coefficients. The faster waves with dipole symmetry (S1) were observed to be preferred in linear models with simple profiles \citep{krause1980meanfield}, and slower waves with quadrupolar symmetry (A1) in more complicated nonlinear models solving also for the dynamics \citep{tuominen2002starspot}. From the viewpoint of dynamo theory, therefore, two migratory active longitudes are an expected result.

Another peculiar property of the dynamo solutions in this regime include their steadiness, i.e. no oscillatory solutions can be found in the very simplest settings \citep[e.g.][]{MTB91}. Therefore, time-dependent phenomena, such as the flip-flop, are difficult to explain in the framework of a simple $\alpha^2$ dynamo mechanism. The usual remedies are to include some differential rotation, that makes time-dependent solutions more easy to excite \citep[e.g.][]{M05}, or to use more complicated profiles for the turbulent transport coefficients \citep[e.g.][]{REO03}, normally working in the same direction. As the observations cannot yet tell whether a polarity change is related to the flip-flop cycle, it is not yet possible to pin down more exactly which type of a dynamo (the possibilities being a pure $\alpha^2$ or a differential rotation aided $\alpha^2\Omega$ dynamo), not to mention which mode of its solutions, is responsible for the observed magnetic fields in rapid rotators.

\section{Conclusions}

We have analysed the active longitudes of the young solar analogue \object{HD 116956} in depth. For this purpose, we have formulated a new analysis algorithm based on Bayesian statistics. This method uses a mixture of two von Mises distributions to model the phase distribution of the active areas manifesting themselves as photometric light curve minima.

Our analysis algorithm allows us to characterise the active longitudes in a systematic way. Both the rotation period $P_{\rm al}$ and the shape and location parameters of the active longitude distribution are parametrised in the model and no subjectivity is involved in their determination. This represents an advance from previously used methods. Typically the rotation period of the active longitudes has been determined with nonparametric methods \citep[e.g.][]{jetsu1996active}. The nonparametric methods can only determine the period, i.e. they do not determine any parameters that characterize the shape or location of the active longitude distribution. Alternatively, the migration rate of the active longitudes with respect to the stellar rotational reference frame can be studied. This has been done by computing linear fits to the phases of the active regions in the reference frame \citep[e.g.][]{berdyugina1998permanent,KBHIST07}. This often involves some subjectivity in choosing which light curve minima correspond to which active longitude, as well as making use of phases well outside of the range $\phi\in[0,1]$.

Our analysis of \object{HD 116956} confirms our previous results \citep{lehtinen2011continuous}, as we find two stable active longitudes with a mutual phase difference of $\Delta\phi\approx0.5$. One of these active longitudes has remained nearly constantly stronger and dominated the overall distribution of active areas on the stellar surface. Our new estimate for the ephemeris of the central meridian passing of the primary active longitude is ${\rm HJD}_{\rm al}=2451177.6135+7.4812E$ (Eq. \ref{ephem}).

We compared the long-term active longitude period $P_{\rm al}=7.8412\pm0.0002$ d with the long-term mean photometric rotation period $P_{\rm rot}=7.817\pm0.003$ d. The estimates for these two periods are markedly different in such a way that the active longitudes lag behind the photospheric rotation. This behaviour may be interpreted as a nonaxisymmetric dynamo mode manifesting itself as a longitudinal dynamo wave propagating in the rotational reference frame of the stellar surface.

It should be noted that the comparison between the periods $P_{\rm al}$ and $P_{\rm rot}$ assumes that the mean photometric rotation period $P_{\rm rot}$ actually measures the stellar rotation period. In the case of weak differential rotation this clearly is so; also theoretical models indicate weak relative differential rotation for rapid rotators. In addition, for synchronously rotating binary systems, which is the case for many \object{RS CVn} stars, we can associate the orbital period $P_{\rm orb}$ with the stellar rotation period. However, even in the case of significant differential rotation, as our previous study suggests for \object{HD 116956}, we are confident in using $P_{\rm rot}$ as the effective measure for the stellar rotation. This is because it tracks the mean rotation period in the latitude zones where the spot activity actually takes place, and where it should be compared with the active longitude rotation period $P_{\rm al}$.

\begin{acknowledgements}
The analysis in this paper was based on observations made as part of the automated astronomy program at Tennessee State University, supported by NASA, NSF, TSU and the State of Tennessee through the Centers of Excellence program. This work has been supported by the Finnish Graduate School in Astronomy and Space Physics and the Academy of Finland project 141017. We thank prof. Karri Muinonen for valuable comments concerning the construction of the MCMC algorithm.
\end{acknowledgements}

\bibliographystyle{aa}
\bibliography{vmcirc}

\end{document}